# Analysis of North Indian Classical Ragas Using Tonnetz

Ananya Giri
National Public School, Indiranagar, Bangalore, India
ananyagiri003@gmail.com

**Abstract:**

In North Indian Classical music, each *raga* has been traditionally associated with a performance time, which supposedly maximises its aesthetic and emotional effects on the listener. The objective of this work was to investigate the structural basis, if any, for the association of *raga*s with different times of the 24-hour span. The *tonnetz* framework has been used to analyze the pitch sets of 65 North Indian Classical *raga*s, and structural similarities have been observed between *raga*s associated with (1) times of transition between day and night, i.e., dawn and dusk, and (2) times between these transitions. These findings could provide some insight into the scientific basis of the age-old *raga*-time relation, and their effects on the perception of the listener.

## 1. Introduction:

Indian classical music (ICM) has two main traditions: North Indian classical music (NICM, known as *Hindustani* music), and South Indian classical music (SICM, known as *Carnatic* music). The two differ in their origins, structure, and lyrical and stylistic aspects. However, both are based on the same fundamental components: the *raga* (राग) and *tala* (ताल).

In NICM, the *tala* is the rhythmic element upon which melodic improvisation is built, and the *raga* is the tonal framework upon which the musician improvises. Similar to the concept of melodic modes in Western Classical Music (WCM), it consists of a definite set of notes (also known as *swaras*), or a scale. Here, the term "pitch set" is used to describe the collection of all notes used in a *raga*. Moreover, each *raga* has a distinctive phrase or sequence of notes, called a *pakad*, and is characterized by specific note progressions and musical motifs that give it a unique mood (*bhaava,* भाव) and emotion (*rasa,* रस).

In ICM an octave is called a *saptak* (सप्तक), and each note is given a name on the basis of the degree on the scale it represents. They correspond to the following WCM names:

| Scale Degree | ICM Name | ICM Abbreviation | WCM Name |
| --- | --- | --- | --- |
| 1st | Shadja | Sa | Tonic |
| 2nd | Rishabha | Re | Supertonic |
| 3rd | Gandhara | Ga | Mediant |
| 4th | Madhyama | Ma | Subdominant |
| 5th | Panchama | Pa | Dominant |
| 6th | Dhaivata | Dha | Submediant |
| 7th | Nishada | Ni | Subtonic/Leading tone |

The pitch sets of *raga*s are subsets of the 12 notes in a chromatic scale, more appropriately referred to as "pitch classes", since in reality the musician covers the hundreds of pitches between notes when transitioning from one to the next in a technique called the *meend* (मींड), best approximated as a glissando. Here, "note" and "pitch class" will be used interchangeably. *Raga* pitch sets have a minimum of 5 notes, and usually include the first, fourth and fifth degrees of the octave (Sa, Ma and Pa). The degrees of the 12 pitch classes on the chromatic scale will be henceforth represented with the following numbers:

**0 1 2 3 4 5 6 7 8 9 10 11**

The adjectives shuddh (शुद्ध, natural), komal (कोमल, flat), and tivra (तीव्र, sharp) are used to denote the variant of the note being referred to. In the case of non-shuddh swaras, lowercase abbreviations are used.

| Degree of Pitch Class on Chromatic Scale | Name of Note | Abbreviation | Equivalent Interval in WCM |
|---|---|---|---|
| 0 | Sa | S | Unison |
| 1 | Komal Re | r | Minor Second |
| 2 | Shuddh Re | R | Major Second |
| 3 | Komal Ga | g | Minor Third |
| 4 | Shuddh Ga | G | Major Third |
| 5 | Shuddh Ma | m | Perfect Fourth |
| 6 | Tivra Ma | M | Augmented Fourth |
| 7 | Pa | P | Perfect Fifth |
| 8 | Komal Dha | d | Minor Sixth |
| 9 | Shuddh Dha | D | Major Sixth |
| 10 | Komal Ni | n | Minor Seventh |
| 11 | Shuddh Ni | N | Major Seventh |

Each *raga* has a prescribed *arohana* (अरोहन; ascending scale) and *avarohana* (अवरोहन; descending scale) across which the pitch set is spread. It also has a *vadi* (वादी) and a *samvadi* (सम्वादी). The *vadi* is the most prominent note in the *raga*, which the musician emphasizes, while the *samvadi* is the second most important note.

There exist thirty-two main parent scales consisting of pitch sets, called *thaat*s (ठाट), under which innumerable NICM *raga*s are classified. Only ten of these *thaat*s are prominent today.

## 1.1 Ragas and Association with Emotions and Time of Day:

*Raga*s are associated with unique moods and emotions. A pioneering study by Mathur et al. (2015) provided empirical support for emotional responses traditionally associated with *raga*s. An important finding demonstrated that major intervals were predictive of reported positive valence, while minor intervals were predictive of reported negative valence. The minor second, in particular, was a significant predictor of a negative valenced emotional response.

*Raga*s have also been traditionally associated with specific seasons and times of the day, which supposedly maximizes their aesthetic and emotional effects on the listener. The 24-hour day is composed of eight subdivisions called *prahar*s (प्रहार), four in the day and four in the night, each lasting three hours. Moreover, the *vadi swara* indicates the time of the day the *raga* is associated with. If the vadi note is part of the *purvanga* part of the *saptak* (Sa, Re, Ga, Ma), the *raga* is meant to be performed between 12pm and 12am. If the vadi is part of the *uttaranga* (Ma, Pa, Dha, Ni) part of the saptak, it is meant to be performed between 12am and 12pm.

However, the *raga*-time connection is debated and has not been studied much. It is not clear whether this is merely something that developed over time, or has a deeper scientific basis.

## 1.2 The Tonnetz:

The tonnetz, a concept used in WCM, is a two-dimensional lattice diagram that represents tonal relationships. First introduced by Euler, it consists of intersecting lines on a Euclidean plane extending infinitely in all directions, each point of intersection representing a note. Line segments joining two notes represent the interval between them. Along the horizontal axis, notes progress in intervals of perfect fifths from left to right. Along the diagonal axes, notes progress in minor third intervals from the top left to the bottom right, and in major third intervals from the bottom left to the top right.

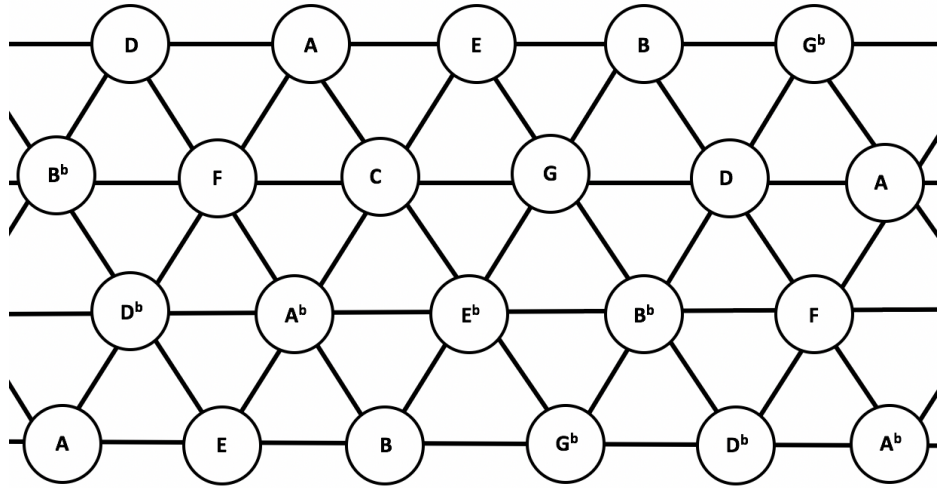

Figure 1: The Tonnetz

Longuet-Higgins and Steedman (1971) have shown that major and minor modes in WCM form clusters with distinct shapes on the harmonic network. Yardi and Chew (2004) applied the tonnetz framework to the pitch sets of NICM *thaats* and proposed connections between their symmetry patterns and the functions of *raga*s, in terms of their associated emotion and times of day. Specifically, the geometry of the structures was used to provide an explanation for the choice of the ten *thaat*s.

However, there is no strict connection between a *raga* and its parent *thaat*; a *raga* that falls under a *thaat* may not include all its notes, and may include other notes as well. Since it is the notes of a *raga* that listeners hear during a performance, this study extends the work of Yardi et al. by using a similar framework to analyze the pitch sets of 65 NICM *raga*s, to explore possible relationships between the pitch sets of *raga*s and their traditionally associated time of day based on the shape of clusters they exhibit on the tonnetz network.

## 2. Method:

This study does a structural analysis of the pitch sets of 65 NICM *raga*s to explore the basis of *raga*-time relation, if any. Following is the list of *raga*s used in this study and their traditionally associated times of the day:

1. **Day**
    1.1. **First Prahar - 6am to 9am:**
        1.1.1. Bhairav
        1.1.2. Ahir Bhairav
        1.1.3. Nat Bhairav
        1.1.4. Jogiya
        1.1.5. Ramkali
        1.1.6. Bilawal
        1.1.7. Alhaiya Bilawal
        1.1.8. Bairagi
    1.2. **Second Prahar - 9am to 12 pm:**
        1.2.1. Deshkar
        1.2.2. Asawari
        1.2.3. Todi
        1.2.4. Basant Mukhari
        1.2.5. Jaunpuri
        1.2.6. Shuddh Sarang
        1.2.7. Bilaskhani Todi
        1.2.8. Charukeshi
        1.2.9. Gurjari Todi
        1.2.10. Sundarkali
        1.2.11. Dev-Gandhar
    1.3. **Third Prahar - 12pm to 3pm:**
        1.3.1. Bhimpalasi
        1.3.2. Patdeep
        1.3.3. Multani
        1.3.4. Pilu
        1.3.5. Bheem
        1.3.6. Dhanashree (Bhairavi Ang)
        1.3.7. Dhanashree (Bhimpalasi Ang)
        1.3.8. Dhani
    1.4. **Fourth Prahar - 3pm to 6pm:**
        1.4.1. Purvi
        1.4.2. Marva
        1.4.3. Shree

## 2. Night

### 2.1. First Prahar - 6pm to 9pm:
   2.1.1. Yaman/Kalyan
   2.1.2. Kedar
   2.1.3. Bhoopali
   2.1.4. Shuddh Kalyan
   2.1.5. Kamod
   2.1.6. Shyam Kalyan

### 2.2. Second Prahar - 9pm to 12am:
   2.2.1. Jog
   2.2.2. Bihag
   2.2.3. Maru Bihag
   2.2.4. Bageshree
   2.2.5. Rageshree
   2.2.6. Chandrakauns
   2.2.7. Puriya
   2.2.8. Abhogi Kanada
   2.2.9. Gorakh Kalyan
   2.2.10. Gunji Kanada
   2.2.11. Bageshree Kanada
   2.2.12. Kafi
   2.2.13. Malhar
   2.2.14. Basant
   2.2.15. Khamaj
   2.2.16. Jaijaivanti
   2.2.17. Durga
   2.2.18. Hansadhwani

### 2.3. Third Prahar - 12am to 3am:
   2.3.1. Malkauns
   2.3.2. Darbari Kanada
   2.3.3. Kafi Kanada
   2.3.4. Hindol
   2.3.5. Jogkauns
   2.3.6. Malgunji
   2.3.7. Shahana Kanada
   2.3.8. Kaushi Kanada

### 2.4. Fourth Prahar - 3am to 6am:
   2.4.1. Bhatiyar
   2.4.2. Lalit
   2.4.3. Sohini

The notes used in each *raga* were mapped on the tonnetz and each pitch set's characteristic cluster shape was obtained. All notes used in the *raga* have been included in the pitch set, irrespective of the differences between *arohana* and *avarohana* notes. Where multiple structures are possible, the most compact structure, i.e., involving the maximum number of connections between notes, has been selected. The 0-7 axis was taken as the frame of reference. Shapes containing the majority of notes above the axis were named "top-heavy" and those containing the majority of notes below the axis were named "bottom-heavy". For example, *raga* Bhairav shows a top-heavy structure (Figure 2.1), and *raga* Multani shows a bottom-heavy structure (Figure 2.3b). The tonnetz mapping for ragas not shown in this paper can be found [here](here).

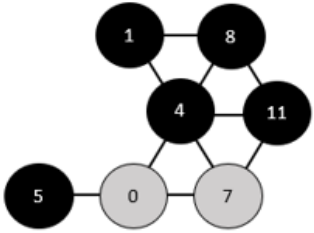
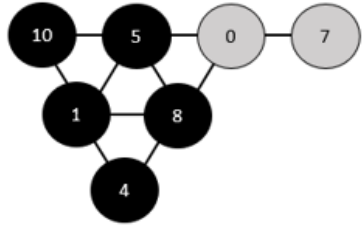
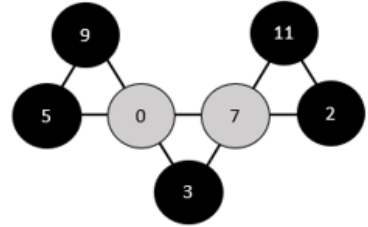
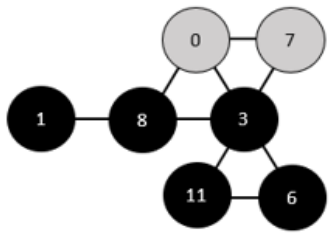
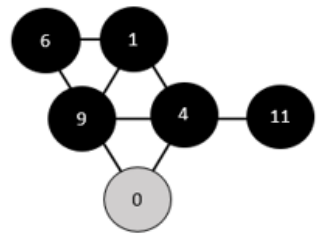

**Figure 2: Daytime Ragas Mapped to Tonnetz**

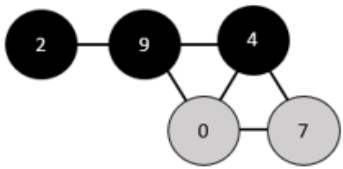
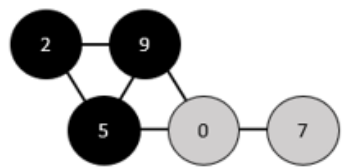
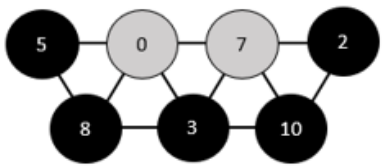
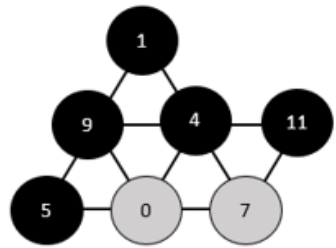

**Figure 3: Night Time Ragas Mapped to Tonnetz**

## 3. Observations:

**Table 1**

| Time | Total Number of Top-Heavy Ragas | Total Number of Bottom-Heavy Ragas | Total Number of Neutral Ragas |
|---|---|---|---|
| Day | 15 | 13 | 2 |
| Night | 21 | 14 | 0 |

**Table 2**

| Prahar of Day | Number of Top-Heavy Ragas | Number of Bottom-Heavy Ragas | Number of Neutral Ragas |
|---|---|---|---|
| 1st (6am-9am) | 6 | 2 | - |
| 2nd (9am-12pm) | 3 | 7 | 1 |
| 3rd (12pm-3pm) | 4 | 4 | - |
| 4th (3pm-6pm) | 2 | 0 | 1 |

**Table 3**

| Prahar of Night | Number of Top-Heavy Ragas | Number of Bottom-Heavy Ragas | Number of Neutral Ragas |
|---|---|---|---|
| 1st (6pm-9pm) | 6 | 0 | - |
| 2nd (9pm-12am) | 11 | 7 | - |
| 3rd (12am-3am) | 2 | 6 | - |
| 4th (3am-6am) | 2 | 1 | - |

1. Two *raga*s (Charukeshi and Purvi) had equally top-heavy and bottom-heavy structures, as a result of which they have been referred to as "neutral" *raga*s. All other *raga*s displayed either top or bottom-heavy structures.
2. It was observed that a striking number of *raga*s associated with the times of transition between day and night (6am-9am and 6pm-9pm) showed top-heavy structures (Tables 2 and 3). Perhaps this is connected to the fact that sunrise and sunset, which are transition times, have been culturally associated with deeper emotions, as innumerable works of literature, music, and art describe dawn and dusk. Moreover, the last prahar of both day and night showed top-heavy structures as well. In total, 3pm-9pm has 8 top-heavy and 0 bottom-heavy *raga*s, and 3am-9am has 8 top-heavy *raga*s and 3 bottom-heavy *raga*s.
3. The times in between transitions slightly tended towards the bottom-heavy structure: the second prahar of day (9am-12pm) showed a bottom-heavy trend (7 bottom-heavy and 3 top-heavy structures), as did the third prahar of night (12am-3am; 6 bottom-heavy and 2 top-heavy structures).

4. It was also observed that the third prahar of the day (12pm-3pm) had an equal number of top and bottom-heavy *raga*s.
5. However, no overall trend was observed linking top-heavy and bottom-heavy structures to daytime and nighttime *raga*s respectively. This is contrary to the work of Yardi et al. (2004).

## 4. Conclusion:

Previous work that examined the structure of *thaat*s on the tonnetz found that *raga*s associated with day tended to be top-heavy and those associated with late night and early morning tended to be bottom-heavy (Yardi et al., 2004). Upon using a similar method of analysis on the tonnetz with the pitch sets of *raga*s, this relation between day and night *raga*s was not found. More specifically, it was found that those *raga*s associated with times of transition between day and night, i.e., 3pm-9pm and 3am-9am, tended to show top-heavy structures. Perhaps this is connected to the fact that sunrise and sunset, which are transition times, have been culturally associated with deeper emotions, as innumerable works of literature, music, and art describe dawn and dusk. Moreover, those *raga*s in between these times of day (specifically, 9am-12pm and 12am-3am) slightly tended towards the bottom-heavy structure on the tonnetz. This correlation between the shape of the pitch sets on the tonnetz and the time of day they are associated with could be helpful in future attempts to explore the *raga*-time relation.

## 5. Future scope:

Though only the pitch sets of *raga*s have been used here, it is important to note that pitch sets alone do not define a *raga*. A *raga* is equally defined by its characteristic musical motifs (*pakad*). For instance, both *raga*s *Miyan ki Malhar* and *Bahar* contain the same notes (S, R, g, M, P, D, n, N) and yet have different flavours because of the way the notes in the scale are combined. In this study, the latter has not been taken into account. Future studies investigating the link between *raga*s and performance time should ideally consider *raga*s' characteristic phrases as well.